\documentclass{aastex63}

\usepackage{ulem}

\usepackage[toc,page]{appendix}
\usepackage{wrapfig}
\received{\today}
\revised{\today}
\submitjournal{TBD}

\shorttitle{SRWL}
\shortauthors{Connor et al.}
\graphicspath{{./}{figures/}}

\begin{document}

\title{Deep Radio Interferometric Imaging with POLISH: DSA-2000 and weak lensing}

\correspondingauthor{Liam Connor}
\email{liam.dean.connor@gmail.com}

\author[0000-0002-7587-6352]{Liam Connor}
\affiliation{Cahill Center for Astronomy and Astrophysics, 
            MC 249-17, California Institute of Technology, Pasadena CA 91125, USA}

\author[0000-0003-0077-4367]{Katherine L. Bouman}
\affiliation{Computing and Mathematical Sciences (CMS),
            Department of Electrical Engineering,, 
            MC 249-17, California Institute of Technology, Pasadena CA 91125, USA}
\affiliation{Cahill Center for Astronomy and Astrophysics, 
            MC 249-17, California Institute of Technology, Pasadena CA 91125, USA}

\author[0000-0002-7252-5485]{Vikram Ravi}
\affiliation{Cahill Center for Astronomy and Astrophysics, 
            MC 249-17, California Institute of Technology, Pasadena CA 91125, USA}
            
\author[0000-0002-7083-4049]{Gregg Hallinan}
\affiliation{Cahill Center for Astronomy and Astrophysics, 
            MC 249-17, California Institute of Technology, Pasadena CA 91125, USA}

\begin{abstract}
Radio interferometry allows astronomers to probe small 
spatial scales that are often inaccessible 
with single-dish instruments. However, recovering the radio sky from an interferometer is an 
ill-posed deconvolution problem that astronomers have worked on for half a century. More challenging still is achieving 
resolution below the array's diffraction limit, known 
as super-resolution imaging. To this end, we have developed a  
new learning-based approach for radio interferometric imaging, leveraging recent advances 
in the classical computer vision problems of single-image super-resolution (SISR) and deconvolution. 
We have developed and trained  
a high dynamic range residual neural network to learn the mapping between 
the dirty image and the true radio sky. We call 
this procedure POLISH, in contrast to the traditional CLEAN algorithm. The feed forward nature of learning-based approaches like POLISH is critical for analyzing data from the upcoming Deep Synoptic Array (DSA-2000). We show that POLISH achieves
super-resolution, and we demonstrate its ability to deconvolve real observations from the Very Large Array (VLA). Super-resolution on DSA-2000 will allow us to measure the shapes and orientations of several hundred million star forming radio galaxies (SFGs), making it a powerful 
cosmological weak lensing survey and probe of dark energy.
We forecast its ability to constrain the
lensing power spectrum, finding that it will be complementary to next-generation optical surveys such as Euclid.


\end{abstract}

\section{Introduction}
\begin{figure}[h!] 
 \centering
  \includegraphics[width=\textwidth]{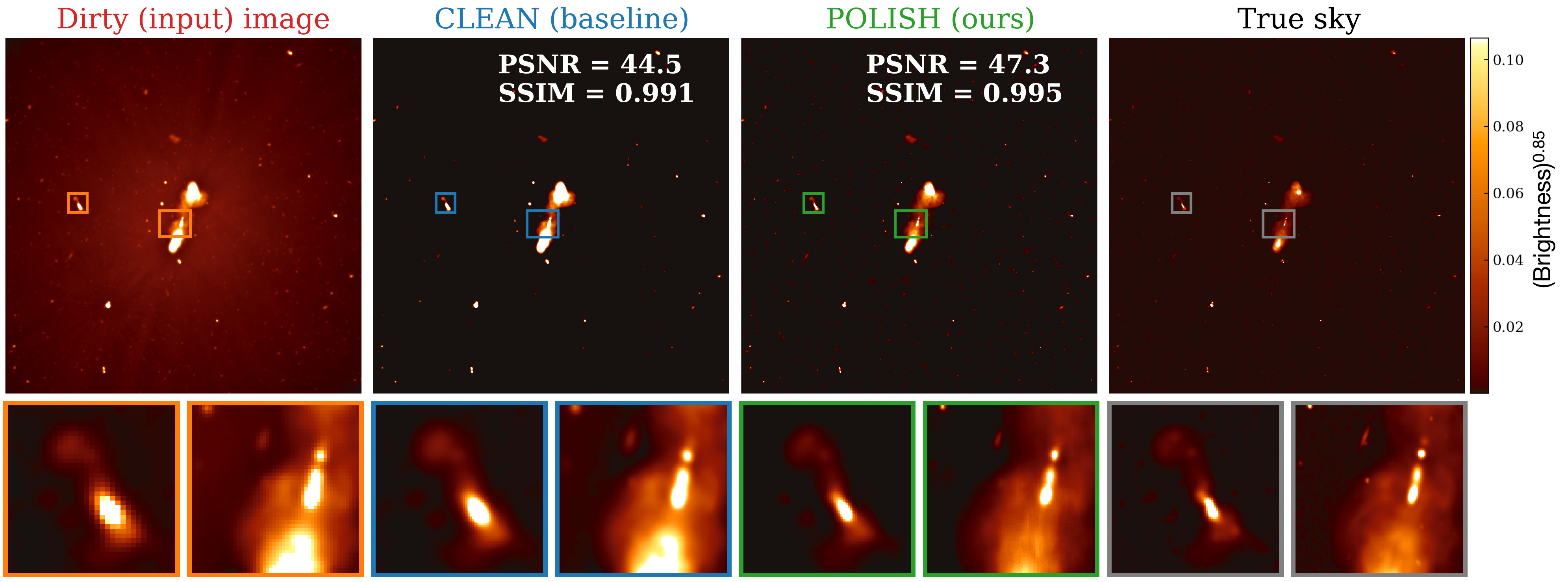}
  \caption{We propose a learning-based approach to recover a super-resolved astronomical image from a ``dirty" input image captured by a radio interferometer. Our method, ``POLISH'' (green), outperforms the image-plane CLEAN algorithm (blue) on realistic simulations of radio images. 
  Training data was significantly different from the radio sky image used in this example (training images were
  composed entirely of simple elliptical Gaussian shapes; see Section~\ref{sec:simulation}), but POLISH is still able to recover structured sources in the image. POLISH resolves finer spatial structures and achieves higher PSNR and SSIM than CLEAN.
  Data are normalized between 0 and 1 then raised to the power of 
  0.85 to better visualize faint structures. Each image 
  is 0.26$\times$0.26\,degrees.}
 \label{fig:ska-fun-lobe}
\end{figure}
High-resolution imaging of astronomical radio sources plays a critical role in studying our Universe. 
However, a fundamental challenge of high resolution imaging is that resolution is limited by aperture size (i.e., dish or 
mirror diameter), rendering single-dish telescopes insufficient in many cases. Alternatively, a radio interferometer is an array of antennas whose radio waves are coherently combined to obtain high resolution images \citep{ryle46}.
Radio interferometry allows one to synthesize a much larger aperture, with the angular resolution now defined by the longest “baseline”, i.e. greatest separation, $D$, between antennas \citep{thompson1988, burke}. The diffraction limit of an interferometric array is given by $\theta\approx\lambda/D$, where $\lambda$ is observing wavelength. Probing scales below this limit is known as 
``super-resolution''.
Interferometry has led to countless new discoveries where single-dish instruments were inadequate, including the first-ever 
direct image of a black hole \citep{eht-2019} and the precise localization 
of extragalactic transients \citep{hallinan-2017, ravi-2019}.

Recovering an image of the sky from an interferometer requires solving an ill-posed deconvolution problem. In particular, the placement of antennas in the interfometer defines a point spread function (PSF) that corrupts the true astronomical image. 
Preeminent observatories such as the 
Karl G. Jansky Very Large Array (VLA) and Atacama Large Millimeter/submillimeter Array (ALMA) have 27\footnote{https://public.nrao.edu/telescopes/vla/} and 66\footnote{https://www.almaobservatory.org/en/home/} radio 
antennas, respectively. 
In contrast, 
the upcoming Deep Synoptic Array in the Big Smoky Valley, Nevada, will contain and combine an unprecedented 2000 antennas \footnote{https://www.deepsynoptic.org}.
The DSA-2000's impressive number of antennas corresponds to a PSF that can be more easily deconvolved, resulting in images with higher fidelity, higher dynamic range, and lower uncertainty. 
This ``radio camera" will be the most powerful radio survey instrument to date and is expected to increase the known sample of astronomical radio sources by a factor of 1000.

The DSA-2000, which will yield unparalleled image quality, will also result in massive computational challenges: the telescope will need to churn through over 80\,Tb/s of data.
Unlike past radio interferometers that can save data to disk for later analysis, it will be impossible for the DSA-2000 to save raw data to disk. Specifically, 
the telescope will not preserve its visibility data. 
It is 
therefore essential that imaging on DSA-2000 be made fast, automated, and deterministic. 
Past methods for radio interferometric imaging are either too slow, require a human-in-the-loop, or result in sub-optimal image quality. 
Thus, it is critical that automated, feed-forward approaches be developed to transform quickly raw dirty images into high-fidelity reconstructions of the radio sky.
We capitalize on recent advances in deep learning to develop an imaging pipeline for the DSA-2000 that can be used to deliver science-ready astronomical images in quasi-real time.

It is also essential that DSA-2000 be able to 
achieve super-resolution. With a PSF width of 
roughly 3\,'', many of the $\sim$\,10$^9$ star forming radio galaxies (SFGs) detected by DSA-2000 will be unresolved 
in the dirty image. If these distant galaxies can be resolved with super-resolution, we will be able to carry out a 
powerful cosmological weak lensing survey on DSA-2000. 
By number of sources, such a survey is competitive with 
upcoming Stage IV optical weak-lensing experiments. Radio weak lensing is also 
highly complementary to upcoming optical surveys because the gravitational lensing signal  
is achromatic, whereas systematics will 
be different across wavelengths \citep{ska-lensing-I, ska-III}. 
We therefore have a strong incentive 
to not only develop a robust feed-forward imaging framework, 
but one that provides super-resolution.

In this work, we demonstrate that recent advances 
in  
the classical problems of single-image super-resolution (SISR) 
and deconvolution through efficient neural network architectures make such imaging with the DSA-2000 possible. 
This manuscript is organized as follows. We first offer background 
on interferometric imaging, the DSA-2000, and the connection to SISR and deconvolution in Section \ref{sec:background}. 
We then describe our deep learning based approach in Section \ref{sec:methods}, 
which we call POLISH (a demonstration is shown in Fig.~\ref{fig:ska-fun-lobe}). 
As we show, POLISH enables imaging at angular resolutions on the sky below 
the intrinsic resolution of the interferometric telescope array 
($\approx$\,3 arcseconds for DSA-2000). 
Results based on simulated DSA-2000 data as well as on real astronomical data from 
the VLA are presented in Section \ref{sec:results}. 
Where previous algorithms were made to 
deconvolve dirty images from sparse arrays, POLISH was designed 
to deconvolve images that are already relatively clean, hence its name. However, our demonstration on VLA data has shown that 
POLISH is a generic radio interferometric deconvolution procedure.
In Section \ref{sec:weaklensing}, we show how super-resolution on DSA-2000 
will enable a powerful weak lensing survey. Finally, we discuss other 
scientific applications and show how feed-forward imaging could 
transform radio astronomy in coming years.

\section{Background}
\label{sec:background}

\begin{figure}
\center
{\includegraphics[width=0.65\textwidth]{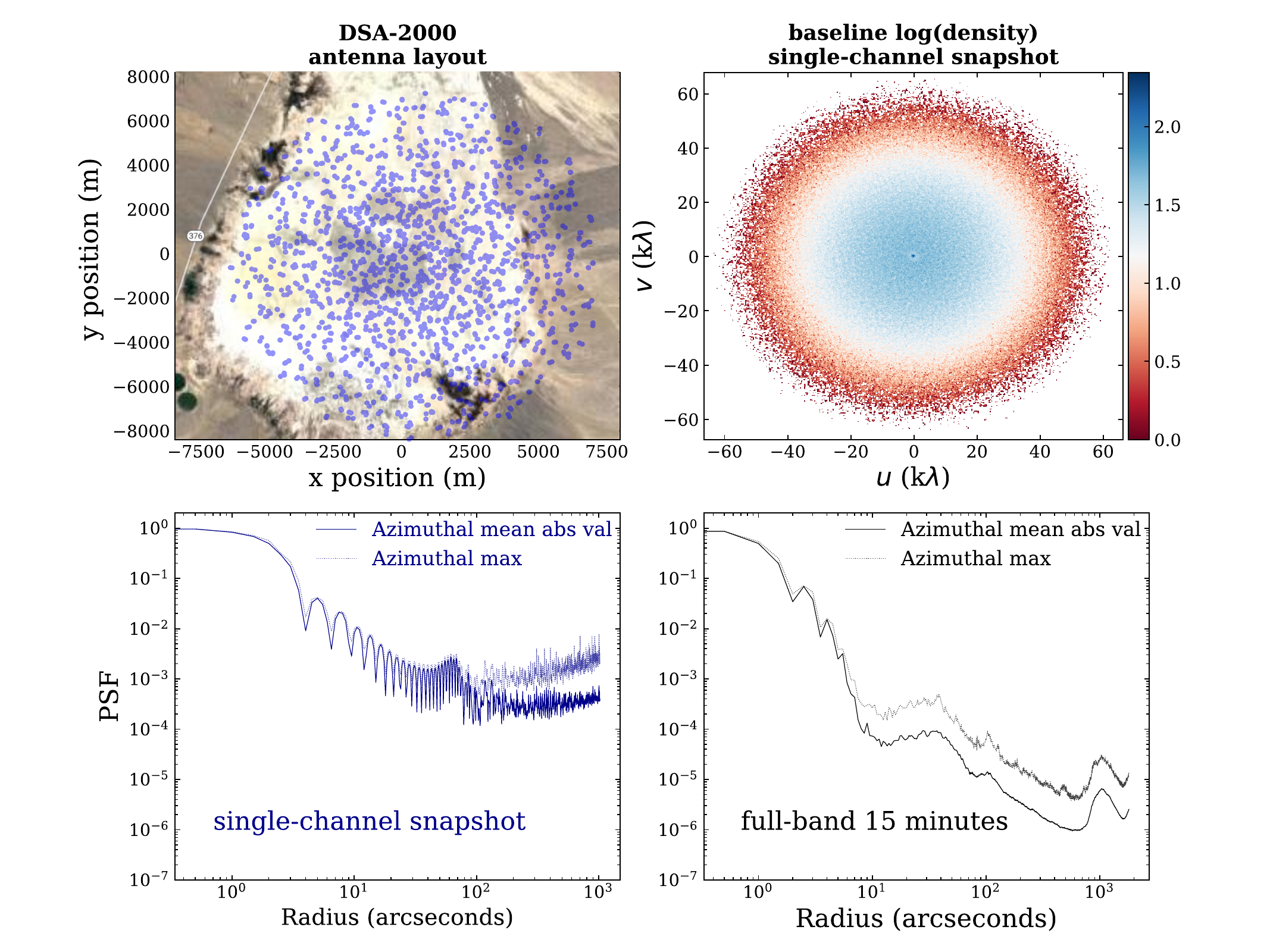}}
{\caption{For our primary training set, 
  we used the DSA-2000 layout shown here. The top left panel shows antenna position assumed 
  for our simulation overlaid on Big Smoky Valley, Nevada. The top right panel shows the array's
  UV coverage for a single frequency at a single time, 
  represented as the logarithm of baseline density. 
  The bottom 
  two panels show an azimuthal average and maximum of the absolute value PSF  
  for a single-channel snapshot (left) and the full band 15-minute integration (right). We expect the 
  DSA-2000 layout to be further optimized for sidelobe suppression and azimuthal smoothness before it is built.}\label{fig:test}}
\end{figure}

\subsection{Interferometric imaging}
\label{sec:imaging}
In radio interferometry, each pair of antennas measures a Fourier coefficient of the underlying astronomical image, where the spatial frequency probed is related to the projected baseline between the antennas as well as the observing wavelength. Imaging then requires solving an inverse problem: recover the true sky image, $I_{\mathrm{sky}}$, using finite, and often sparse, spatial frequency measurements.
Given enough antennas, one could densely sample the entire spatial frequency plane. In the absence of noise, these complex measurements, known as ``visibilities'' $V$, could then be used to recover an image through an inverse Fourier transform:
\begin{equation}
    I_{\mathrm{sky}} = \mathcal{F}^{-1}\left (V \right ).
    \label{eq:sky}
\end{equation}
Since we only have access to a finite number of spatial frequency Fourier coefficients 
in practice (i.e. a finite number of antennas pairs and finite range of radio frequencies), we do not have full access to $V$. We mathematically represent this by applying a sampling function, $S$, to the noisy measurements in the Fourier plane,
\begin{equation}
     I_{\mathrm{d}} = \mathcal{F}^{-1} \left((V+ n_{\rm vis})\,S \right),
    \label{eq:sky2}
\end{equation}
where $n_{\rm vis}$ represents independent Gaussian noise at each spatial frequency and 
$I_{\mathrm{d}}$ is known as the ``dirty image''. Invoking the convolution theorem, we can rewrite this equation as 
\begin{equation}
    I_{\mathrm{d}} = \mathcal{F}^{-1}\left (V \right ) 
    \otimes \,\mathcal{F}^{-1} \left (S\right) + n_{\rm sky},
\end{equation}
where $\otimes$ denotes convolution and $n_{\rm sky} = \mathcal{F}^{-1}\left( n_{\rm vis} \right ) \otimes \,\mathcal{F}^{-1} \left (S\right)$.
The inverse Fourier transform of the sampled visibilities, $I_{\mathrm{d}}$, is degraded 
by the sampling function.
By noting that the Fourier transform of our sampling function
is just the instrument's PSF, we 
can replace $\mathcal{F}^{-1} \left (S\right)$ with a convolution kernel 
$k$. This kernel depends not only on the telescope configuration but also the radio bandwidth, as 
can be seen in Figure \ref{fig:NNarchitecture}. Finally, we are free to choose a pixelization scheme such that the resolution of the recovered image, $\hat{I}_{\mathrm{sky}}$, is $s$ times higher resolution than 
$I_{\mathrm{d}}$. By substituting Eq.~\ref{eq:sky}, we obtain a mathematical description of the noisy and low-resolution image obtained by the interferometer, $I_{\mathrm{d}}$, as a function of our target, $I_{\mathrm{sky}}$:
\begin{equation}
    I_{\mathrm{d}} = \left ( I_{\mathrm{sky}} \otimes \,k \right )\downarrow_s + n_{\rm sky},
\label{eq:sisr-sky}
\end{equation}

\noindent where $\downarrow_s$ is a 
downsampling operation. The inversion 
is challenging because for every $I_d$ there exist 
an infinite number of $I_{\mathrm{sky}}$ that would satisfy 
Eq.~\ref{eq:sisr-sky}: i.e., interferometric deconvolution is an ill-posed inverse problem. It is also difficult for computational reasons, as the large spatial extent of the PSF, or kernel $k$, require large input and output images.

The goal of our work is to learn this mapping such that 
Eq.~\ref{eq:sisr-sky} can be inverted and an 
estimate for the true sky, $\hat{I}_{\mathrm{sky}}$, can 
be automatically recovered with a single pass through a neural network. 

\subsection{The Deep Synoptic Array (DSA-2000)}

The Deep Synoptic Array, i.e., DSA-2000, is an ambitious new instrument expected to 
be complete by 2026 \citep{dsa-2000-whitepaper}. It will survey the radio universe with unprecedented speed and depth, 
but this power comes at the expense of serious computational challenges. The DSA-2000 will consist of 2000$\times$5\,m dual-polarization antennas observing the sky between 0.7 and 2.0\,GHz. By the Nyquist–Shannon sampling theorem, the DSA-2000 must
sample the electric field at more than 4 billion times 
per second, resulting in a raw data rate of over 10\,TB/s. At the next stage in the pipeline, a single image of the sky produced using 15-minutes of visibility measurements requires
$\approx$\,20\,TB of data. This tremendous data rate prevents the use of imaging algorithms that store data and iteratively transform from visibilities in the Fourier plane to images in the sky plane. Instead, computational resources require that the DSA-2000 imaging pipeline rapidly recover an image of $I_{\mathrm{sky}}$ from $I_{\mathrm{d}}$ directly. 
In our proposed method, POLISH, we have developed a feed forward approach to invert a 2,000$\times$2,000 pixel image,
$I_{\mathrm{d}}$, into a sky 
reconstruction, $\hat{I}_{\mathrm{sky}}$, in just a few seconds on a laptop. 

\subsection{Prior interferometric imaging techniques}
Over the past half-century, considerable effort has 
been put into radio interferometric image reconstruction. 
While significant progress has been made, state-of-the-art algorithms 
do not satisfy the requirements for efficient super-resolution 
imaging on the DSA-2000. Imaging algorithms can be broadly categorized into two
methodologies: CLEAN deconvolution and regularized maximum likelihood (RML) approaches.
\vspace{0.1in}

\noindent {\bf CLEAN   } The de facto standard image reconstruction algorithm used in radio astronomy 
is known as CLEAN \citep{hogbom, clark}. 
CLEAN attempts to iteratively 
deconvolve the PSF kernel $k$ from ${I}_{\mathrm{d}}$ to recover $I_{\mathrm{sky}}$,
by assuming the sky is made up of a collection of point-sources. At each step, single pixels are added to a sky model
at the locations of bright pixels in ${I}_{\mathrm{d}}$. The model pixels are scaled by the brightness of the corresponding pixels in ${I}_{\mathrm{d}}$.
The sky model is then 
convolved with a Gaussian ``restoring 
beam'' to provide a smooth estimate of the recovered image.
This final blurring step, necessitated by the point-source assumption in CLEAN, precludes CLEAN from recovering
high spatial frequencies. This is because the restoring beam 
smooths data to the angular scale of the central 
component of the PSF. Besides not being able to deliver super-resolution, standard iterative algorithms like CLEAN introduce severe imaging artifacts near very bright or diffuse radio sources (see Fig.~\ref{fig:vla}). There are variants 
on the standard H{\"o}gbom CLEAN algorithm that allow 
for better reconstruction of diffuse or structured images, 
such as multi-scale CLEAN that are less sensitive to the point-source assumption \citep{wakker-1988, cornwell2008}. In this paper we focus on standard image-plane CLEAN, as that is the likely alternative 
algorithm to POLISH that would be used on a 
radio camera such as DSA-2000.
%
\vspace{0.1in}

\noindent {\bf Regularized Maximum Likelihood  } Regularized Maximum Likelihood (RML) methods are Bayesian inspired techniques that solve for an image by minimizing an objective function, which includes a term that penalizes poor data fit and an image regularization term \citep{wiaux2009, Bouman-2016, sun-2020}. Some image regularization terms employed include maximum entropy \citep{Cornwell-1985, mem86}, total variation, and $\ell_1$ sparsity 
\citep{pratley-2018}.
RML methods have had recent success in the radio interferometry community in their use for reconstructing the first image of a black hole, M87* 
\citep{eht-2019}. Since RML methods do not rely on the CLEAN assumption that the sky is composed of 
point-sources, they can achieve super-resolution \citep{Abdulaziz-2019}. However, they are very computationally expensive since they are iterative and model every measurement in the visibility domain.
Though RML methods provide a more natural 
way of achieving super-resolution than CLEAN, these algorithms 
are not well suited for the DSA-2000 
where fast, feed-forward image reconstruction is essential. 

\subsection{Deconvolution \& single image super-resolution}
\label{sisr}

Deconvolution is a classic inverse problem that aims to reconstruct a sharp image by removing an imaging system's PSF from its measured blurry image 
\citep{deconv-1986,deconv-astro}. Traditional approaches to deconvolution include Weiner filtering and regularized inversion, however recently deep learning approaches have demonstrated impressive results on the task \citep{xu2014deep, yan-2016}. Single image super-resolution (SISR) is an ill-posed inverse problem in which one attempts to obtain a
high-resolution (HR) output from its corresponding low-resolution (LR) 
image \citep{sisr-2009, sisr-2014}. Here too, deep learning based approaches to SISR have made significant progress, and efficient neural network architectures 
continue to be developed to learn the mapping between the 
high- and low-resolution images \citep{srcnn, srgan, sisr-review}.
SISR shares conceptual similarities with deconvolution, since one can frame the SISR problem as removing the downsampling operation's PSF from the image. However, unlike deconvolution, the mismatch in dimensions between the desired HR and input LR image causes the inverse problem to be ill-posed. The challenge of image reconstruction in radio interferometry can be formulated as a joint deconvolution and SISR inverse problem. In fact, replacing $I_{\mathrm{d}}/I_{\mathrm{sky}}$ with 
$I_{\mathrm{LR}}/I_{\mathrm{HR}}$ in Eq.~\ref{eq:sisr-sky} 
reproduces the generalized SISR image degradation equation used in \citep{sisr-review}.

There are nonetheless four critical differences between standard deconvolution-SSIR and astronomical imaging with radio interferometers. 
First, images of the radio sky require large dynamic ranges. In the annual New Trends in Image Restoration and Enhancement (NTIRE) data challenge for SISR, the image pairs were RGB 8-bit precision and roughly 2,000$\times$2,000 pixels. A typical 1\,$\deg^2$ field on the sky measured with the DSA-2000 radio interferometer will have $\sim$\,20,000 unique sources with a range of brightness values approaching 
five orders of magnitude. Thus, we require at least 16-bit integers 
or preferably, 32-bit floats. 
Second, both the sky and instrument response are dependent on observing 
frequency. We therefore prefer multifrequency input data with an arbitrary 
number of channels, a generalization of the three color channels used for RGB images. 
Third, the PSF in radio astronomy can be highly spatially extended, unlike the simple Gaussian blurring kernel or bi-cubic downsampling typically assumed in many super-resolution applications. The PSF 
can also have significant radial and azimuthal structure (refer to Fig.~\ref{fig:NNarchitecture}). Fourth, while the response of our instrument is 
knowable in principle, calibration errors and other systematics lead to perturbations of the PSF. Therefore, our 
neural network must learn not just a single kernel, 
but a physically realistic distribution of PSFs (akin to partially blind deconvolution).

\begin{figure}[tb] 
 \centering
  \includegraphics[width=0.98\textwidth]{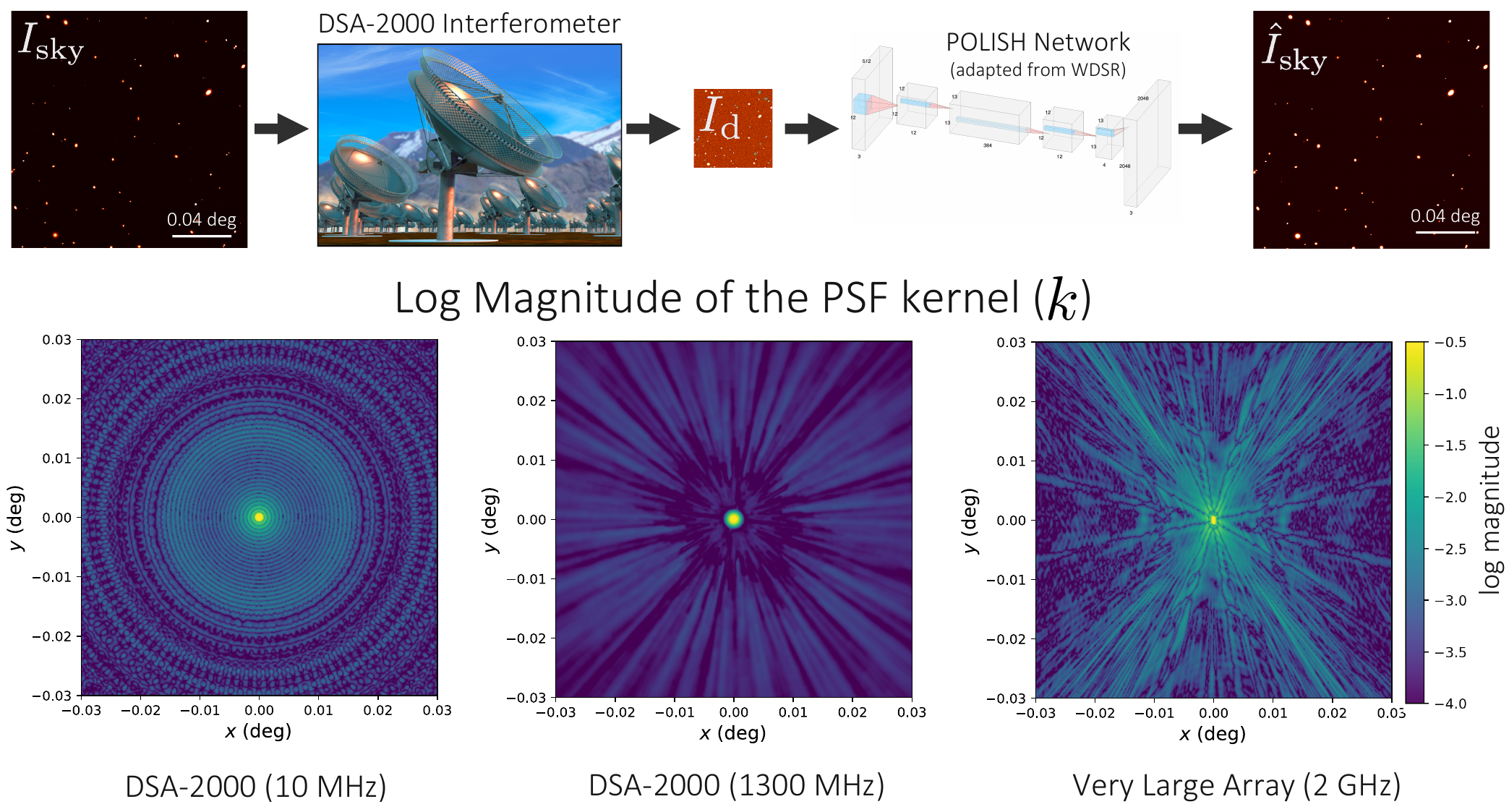}
  \caption{A schematic of the POLISH method of 
  image reconstruction. The top row shows how the true sky 
  is transformed by a radio interferometer and then recovered by our neural network.
  The bottom three figures show the PSF kernels used in this work: the DSA-2000 with just 10\,MHz 
  of its bandwidth used (left), and the full-band (1300 MHz) PSF 
  of the DSA-2000 (middle), and an example VLA array configuration (right). Note that these kernels are plotted over the same extents as the observed sky images, highlighting that the kernels cause artifacts from bright sources that extend over the entire images.
 }
 \label{fig:polish-diagram}
\end{figure}

\section{Methods}
\label{sec:methods}
The application of neural networks to super-resolution and 
deconvolution aims to learn the mapping between an input 
(convolved, lower-resolution) image and the true, higher-resolution image. This is done by training the 
neural network with a large number of image pairs ($I_d$/$I_{\mathrm{sky}}$ in our case) such that, 
when fed an input image the network can reproduce an 
accurate reconstruction of $I_{\mathrm{sky}}$. The training process penalizes any disagreement between the 
reconstructed image $\hat{I}_{\mathrm{sky}}$ and 
the true image, $I_{\mathrm{sky}}$, through its 
``loss function''. In effect, the neural network 
learns deconvolution. Radio interferometric imaging 
is therefore a natural application of the striking recent 
advances in learning-based deconvolution and super-resolution.

\begin{figure}[htbp] 
 \centering
  \includegraphics[width=0.7\textwidth]{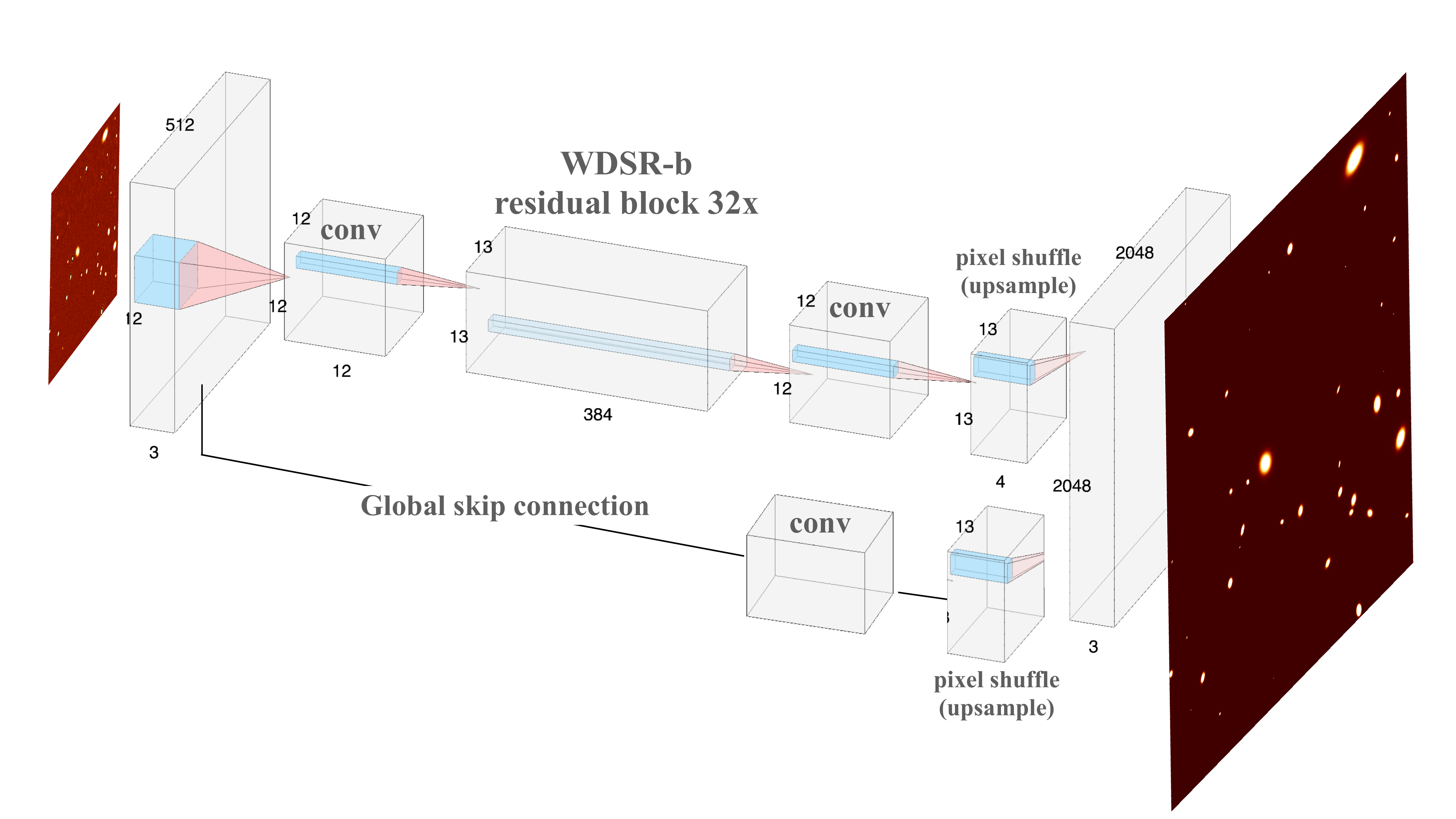}
  \caption{\small The neural 
  network architecture that we have adapted 
  for radio imaging, based on WDSR \citep{wdsr}. 
  A dirty image is fed 
  to the network as an input and its output is a 
  deconvolved reconstruction of the sky with 
  $s$ times greater resolution in each direction. 
  The network relies on residual 
  blocks and a global skip connection. Batch normalization is removed.}
 \label{fig:NNarchitecture}
\end{figure}

We modify the previously proposed Wide Activation for Efficient and Accurate Image 
Super-Resolution (WDSR) \citep{wdsr} architecture to allow for 
high-dynamic range and multifrequency input/output data, as well as to 
handle uncertainty in the PSF. 
This architecture is shown in 
Fig.~\ref{fig:NNarchitecture}.\footnote{We based our implementation on the publicly available code at https://github.com/krasserm/super-resolution under the Apache License 2.0 license.}
Our adaptation, POLISH, takes 
as input a dirty image, $I_{\mathrm{d}}$: a 
tensor with dimension ($n_x$, $n_y$, $n_f$) where $n_x$ and $n_y$ are spatial dimensions and 
$n_f$ is the number of frequency channels. POLISH outputs a 
super-resolved reconstruction of the sky, $\hat{I}_{\mathrm{sky}}$, 
whose shape is ($s\,n_x$, $s\,n_y$, $n_f$). Here, 
$s$ is the upsampling factor and takes values 2 or 3
in our work. The architecture depends on a residual network design with a 
single global skip connection and a residual block 
called ``WDSR\_b''. The final operation before the global skip connection and the primary 
path are summed is known as a pixel shuffle. These layers upsample the image by a factor of $s$, using a sub-pixel convolution 
with stride $1/s$ \citep{pixel-shuffle}.
To ensure high dynamic range deconvolution, POLISH now accepts as input 16-bit, or 32-bit 
data. 

We train the POLISH network, $G_{\theta}$, on $N$ image pairs
$ \left(I_{\mathrm{d}}^{(n)} , I_{\mathrm{sky}}^{(n)} \right)$, related through Eq.~\ref{eq:sisr-sky}, to minimize an $\ell_1$ loss: 
\begin{equation}
    \hat{\theta} = \arg \min_{\theta} \sum_{n} \left|\left| I_{\mathrm{sky}}^{(n)} - G_{\theta} \left(I_{\mathrm{d}}^{(n)} \right)\right|\right|_1
\end{equation}
The model is trained with roughly 
$10^6$ total epochs and a constantly decaying learning 
rate between $10^{-3}$ to $5\times10^{-4}$. 
After each 
1000 epochs of training, the model is tested on 10 images 
from the validation set, and a checkpoint is saved if its 
PSNR has improved. 
For results presented in this main paper, we use 800 image pairs for our training data and 100 pairs in our validation set.
Each model takes roughly 20 GPU hours to train using an NVIDIA TITAN RTX GPU with 
24\,GB of memory\footnote{Training was 
done on an internal cluster. The total compute spent on this project was 
roughly 300\,GPU hours.}. Before PSNR is calculated, or 
an $I_{\rm d}$ is fed to POLISH,
data are normalized such that the image's pixel 
values fall between 0 and 2$^{n_{\rm bit}}$-1. 

\subsection{Learning a PSF distribution}
\label{sect:PSF}
In radio interferometry an idealized PSF kernel is known {\it a priori}, 
as it is simply the inverse Fourier transform of 
the sampling function, $k_{\mathrm{ideal}} = \mathcal{F}^{-1}(S)$.
However,
in practice the instrumental response will cause the PSF to deviate from the ideal model
due to, e.g., calibration errors. 
We therefore want the POLISH network to learn a distribution of PSFs, 
rather than a single PSF kernel. We build this into our training 
set by randomly perturbing the PSF 
in physically motivated ways when generating training image pairs. 
In particular, for each training pair the dirty image 
is created by applying a kernel that has been transformed 
in the following way,
\begin{equation}
    k = \Lambda_3(\phi) \Lambda_2(\gamma) \Lambda_1(\alpha)\,k_{\mathrm{ideal}}.
    \label{eq:psf-dist}
\end{equation}
These three transformations include a radial stretch, $\Lambda_1(\alpha): r\rightarrow \alpha\,r$
where the PSF shape is conserved but 
its size varies by tens of percent; this degradation is motivated by
ionospheric effects that broaden the kernel but never shrinks it (i.e., $\alpha$ is drawn from 
(0,1])) \citep{ionosphere}. $\Lambda_2(\gamma)$ warps the PSF with random distortions \citep{psf-distort}, characterized 
by a parameter $\gamma$, as seen in Fig~\ref{fig:psf-pertubation}.
Finally, an affine transformation\footnote{https://mathworld.wolfram.com/AffineTransformation.html} $\Lambda_3(\phi)$ both contorts the PSF and a shifts it with a  
spatial translation, similar to what occurs when a telescope is mispointed.

\begin{figure}[tb] 
 \centering
  \includegraphics[width=\textwidth]{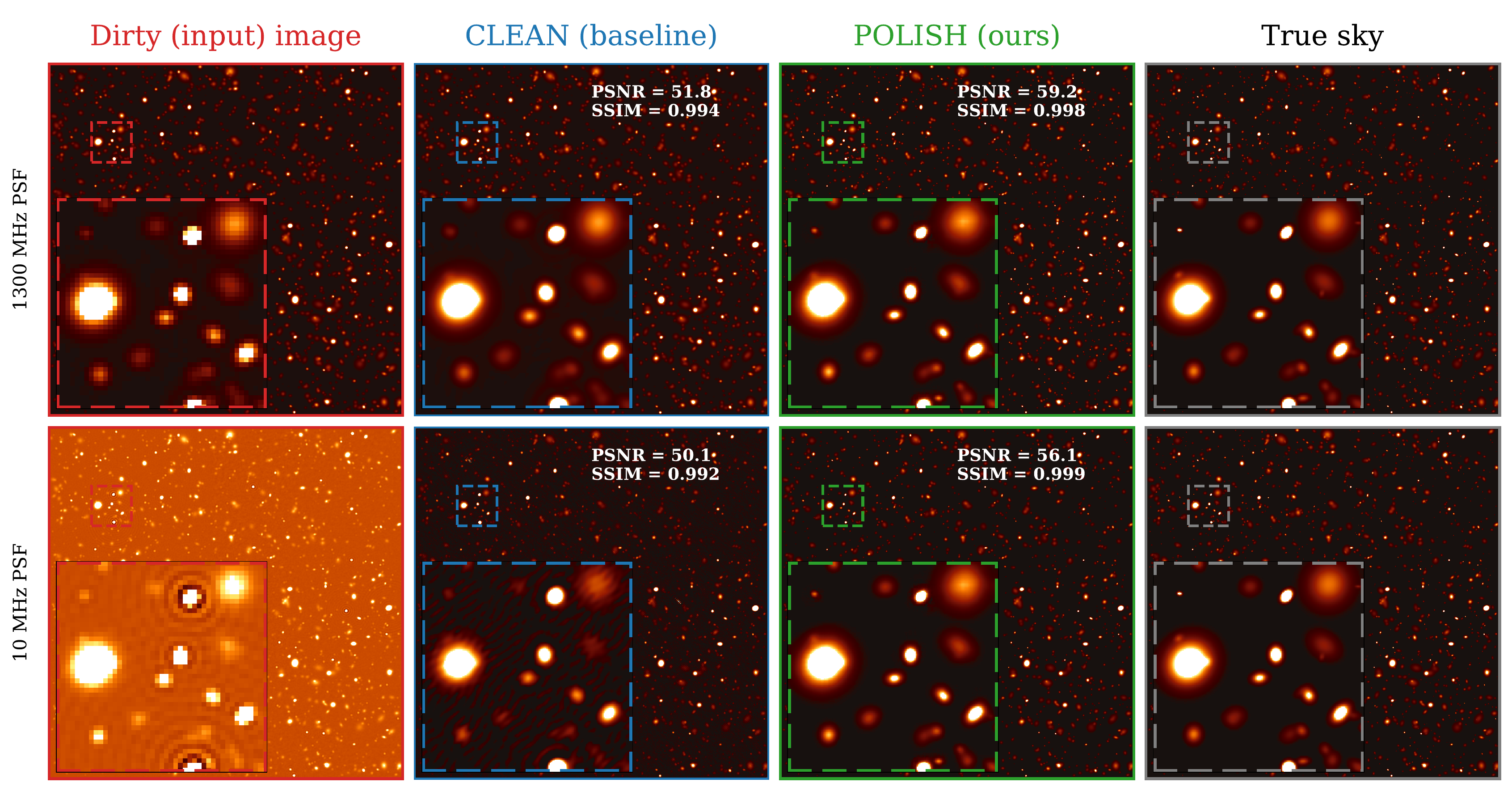}
  \caption{A simulated DSA-2000 interferometric observation of distant 
  radio galaxies and the image reconstruction by image-plane CLEAN (middle left)
  and POLISH (middle right). The bottom 
  row shows the sky convolved with DSA-2000's PSF using less than 
  1$\%$ (single-channel snapshot); this case is critical for HI and OH science. 
  The top row PSF uses the 
  full-band with 15-minutes of integration and has a more complete sampling in the Fourier plane. In both cases POLISH outperforms the image-plane CLEAN baseline, especially for the 10\,MHz PSF where substantial artifacts exist in the CLEAN reconstruction. The brightness of each image is independently scaled for visualization purposes.}
 \label{fig:galaxyfig}
\end{figure}




\subsection{Loss function and performance metric}
We have tried training our model using  
$L^1$ and $L^2$ loss functions. Both 
are pixel-wise loss functions that measure the mean absolute error and 
mean squared error between input and output images, respectively. We find that the $L^1$ loss gives better results in terms of time for training to converge, in line 
with previous EDSR and WDSR networks \citep{edsr, wdsr}.
The final results presented here were trained using $L^1$ 
loss. 

As a way of assessing the veracity of our reconstructed image
and the performance of our neural network, 
we use two metrics. The first is peak signal-to-noise ratio ($PSNR$), 
which a standard statistic used in computer vision. 
PSNR is a pixel-by-pixel estimator of the ratio of 
maximum signal and corrupting image noise, expressed in decibels. It is given by,

\begin{equation}
    PSNR = 20 \log{\left ( \frac{MAX_I}{MSE} \right )},
\end{equation}

\noindent where $MAX_I$ is the maximum possible pixel value 
in an image, and $MSE$ is the mean squared error across all 
pixels. For an 8-bit RGB image, $MAX_I$ is 255. In general, it is $2^{N_{\rm b}}-1$ where $N_{\rm b}$ is the number of bits per sample. The $MSE$ is calculated as the average square difference between the true image and the super-resolved image ($I_{\rm SR}$), given by,

\begin{equation}
    MSE = \frac{1}{N_l\,N_m} \sum_{l} \sum_{m} \left ( I_{\rm sky}(l, m) - I_{\rm SR}(l, m)\right )^2.
\end{equation}

$PSNR$ depends on pixel-wise error, making it sensitive to 
translations of the image, or ``astrometric error'' in our case. Since we are concerned here with small distortions to galaxy shapes, we also employ a more 
perception-based metric known as the structural similarity index measure ($SSIM$)\footnote{https://en.wikipedia.org/wiki/Structural\_similarity} \citep{ssim}. $SSIM$ compares various windows within 
an image to asses the statistics of those regions, rather 
than directly computing absolute differences in pixels. For two 
boxes $x$ and $y$ with dimensions $N\times N$,

\begin{equation}
    SSIM(x,y) = \frac{(2\mu_x\mu_y + c_1)(2\sigma_{xy} + c_2)}{(\mu_x^2 + \mu_y^2 + c_1)(\sigma_x^2 + \sigma_y^2 + c_2)}
\end{equation}

\noindent Here $\mu$ is the mean of that box, $\sigma^2$ is 
variance, $\sigma_{xy}$ is covariance, and $c_1$ and $c_2$ are two variables to stabilize the division with weak denominator. We use both $SSIM$ and $PSNR$ to test the performance of POLISH.

\section{Experimental setup \& results}
\label{sec:results}

The performance of POLISH is assessed on both realistic simulations mimicking data from the DSA-2000, as well as real experimental data from the VLA at 2--4\,GHz. We compare the 
POLISH reconstruction with image-plane CLEAN and 
the true image (in the case of simulated sky data), using both
PSNR and SSIM to quantify reconstruction quality. 
These are summarized in Table~\ref{tab:results}.

\subsection{Synthetic training data}
\label{sec:simulation}

We model the radio sky and measurement noise to simulate a physically realistic training set. This is done by first simulating a $\sim$2,000$\times$2,000 pixel image of the true sky, $I_{\mathrm{sky}}$ with a 
field of view of $\sim$0.3$\times$0.3\,deg. DSA-2000 will primarily detect radio galaxies 
with a detection rate corresponding to 
an average of 13,000 galaxies per square degree above a signal-to-noise (SNR) of 6 \citep{dsa-2000-whitepaper}. 
We populate each 
realization of $I_{\mathrm{sky}}$ with $N$ sources, where 
$N$ is randomly drawn from a Poisson distribution whose mean is 13,000 times the sky coverage in degrees. The 
galaxies are 2D Gaussian ellipsoids with random orientation 
angles on the sky. The statistics of their brightness, 
ellipticity, and angular size are chosen to match the empirically 
measured distributions for radio galaxies \citep{tunbridge-2016}. The galaxy peak flux values range from a few to several thousand $\mu$Jy/pixel. Finally, we model the spectral index of each source with a Gaussian distribution 
centered on -0.55 and with standard deviation 0.25, based 
on the observed spectra of SFGs near the 
DSA-2000 band, 0.7--2\,GHz \citep{spec_ind_sfrg}.

A dirty image, $I_{\mathrm{d}}$, corresponding to each true sky image is obtained through the process mathematically described in Eq.~\ref{eq:sisr-sky}. The PSF kernel used to degrade $I_{\mathrm{sky}}$ is obtained by sampling from Eq.~\ref{eq:psf-dist}. The unperturbed PSF is computed for each telescope configuration using the packages {\tt casa} \citep{casa} and {\tt wsclean} \citep{wsclean}. Before applying the PSF kernel to obtain $I_{\mathrm{d}}$, zero-mean {\it i.i.d.} Gaussian noise with a standard deviation $\sigma_{\mathrm{sky}}$ is applied to mimic $n_{\mathrm{sky}}$. The 
convolved image is then downsampled by a factor $s$.
The parameters $\sigma_{\mathrm{sky}}=0.5$ $\mu$Jy/pixel and $s=3$ are used to simulate DSA-2000 dirty images, while values of $\sigma_{\mathrm{sky}}=3$ $\mu$Jy/pixel and $s=2$ are used for generating VLA training data used in Section \ref{sec:vla}. 

Results presented in the main paper consist of 16-bit single-channel input data, where a multifrequency dependence has been included in the PSF kernel but not $I_{\mathrm{sky}}$. On DSA-2000, we 
use two PSFs: one that uses the full radio bandwidth (1300\,MHz) with 
15 minutes of earth-rotation synthesis and one 
that uses just \,$\sim$\,1$\%$ of 
the observing bandwidth (a narrow 10\,MHz range of wavelengths) for a single 
snapshot, shown in Fig.~\ref{fig:polish-diagram}. 
If the PSF is generated across multiple frequencies, it is effectively averaged over the band to produce a single channel.

\subsection{Results}

Figure \ref{fig:ska-fun-lobe} visually compares the performance of POLISH against the baseline CLEAN algorithm on synthetic DSA-2000 observations, 
using an SKA Data Challenge image as $I_{\rm sky}$ \citep{ska-data-challenge}. 
The input dirty image, corresponding to a 1300 MHz bandwidth PSF, is shown on the left.
Although POLISH was only trained on images with Gaussian sources, it is able to recover more complex underlying sharp structure. As demonstrated by the increase in PSNR and SSIM, POLISH recovers a substantially improved image compared to both the measured dirty image and the CLEAN reconstruction, at nearly 3\,dB higher than the latter. Image-plane CLEAN produces more speckle and sidelobe artifacts than 
POLISH, leading to false-positive detections of 
radio galaxies in our simulation. For example, 
\texttt{Source-Extractor} tends 
to find approximately one false-positive galaxy 
per twenty simulated galaxies in the CLEAN 
reconstruction, whereas the POLISH neural 
network ``hallucinates'' just one source per fifty true galaxies.

Figure \ref{fig:galaxyfig} shows a visual example of POLISH results on DSA-2000 synthetic data, comparing results on data generated with a 10 MHz and 1300 MHz bandwidth PSF. 
In order to best see the improvement, each panel shows a zoom-in on a 0.06\,$\deg\times$0.06\,$\deg$ region of the full simulated sky, where dozens of radio galaxies 
with different sizes, shapes, and orientations can be seen. 
In the 10 MHz bandwidth PSF, less than $1\%$ of the available 
radio wavelengths are used to construct the PSF; reconstructing an image with only a small portion of the full PSF bandwidth is important for HI and OH studies on DSA-2000. This region of the sky has two particularly bright 
``galaxies", which makes the PSF visible in the 10 MHz bandwidth dirty image.

\begin{table}[]
\center
\begin{tabular}{lllll}
\hline
Bandwidth (MHz)          & $PSNR$ (POLISH)                    &  $PSNR$ (CLEAN)                    &  $SSIM$ POLISH                  &  $SSIM$ CLEAN                       \\ \hline
\multicolumn{1}{c}{1300} & \multicolumn{1}{c}{55.9$\pm4.7$} & \multicolumn{1}{c}{50.0$\pm6.0$} & \multicolumn{1}{c}{0.998$\pm$0.0016} & \multicolumn{1}{c}{0.989$\pm$0.007} \\
\multicolumn{1}{c}{10}   & \multicolumn{1}{c}{55.1$\pm3.8$} & \multicolumn{1}{c}{47.4$\pm3.6$} & \multicolumn{1}{c}{0.988$\pm$0.0016} & \multicolumn{1}{c}{0.976$\pm$0.009} \\ \hline
                         &                                  &                                  &                                      &                                    
\end{tabular}
\caption{Results from image reconstruction 
comparison between CLEAN and POLISH for 
the full-band case (1300\,MHz) and 
the single-channel scenario (10\,MHz). 
POLISH achieves a higher Peak Signal-to-Noise Ratio (PSNR) 
and structural similarity index measure (SSIM) than CLEAN, 
indicating better pixel-wise and perceptual 
fidelity, respectively.}
\label{tab:results}
\end{table}

In both cases POLISH substantially outperforms the CLEAN baseline. POLISH achieves a mean 
PSNR$=$55.9$\pm4.7$ and SSIM$=$0.998$\pm$0.0016 for our 
radio galaxy simulation with the 
1300\,MHz PSF on 50 validation images, 
compared with PSNR$=$50.0$\pm6.0$ and SSIM$=$0.989$\pm$0.007 on CLEAN.\footnote{The number of validation images was limited by CLEAN computational requirements.} The 10\,MHz PSF results in PSNR$=$55.1$\pm3.8$ and SSIM$=$0.988$\pm$0.0016 using POLISH. 
For CLEAN this was PSNR$=$47.4$\pm3.6$ and SSIM$=$0.976$\pm$0.009. These results are 
presented in Table~\ref{tab:results}.


\begin{figure}[tb] 
 \centering
  \includegraphics[width=0.85\textwidth]{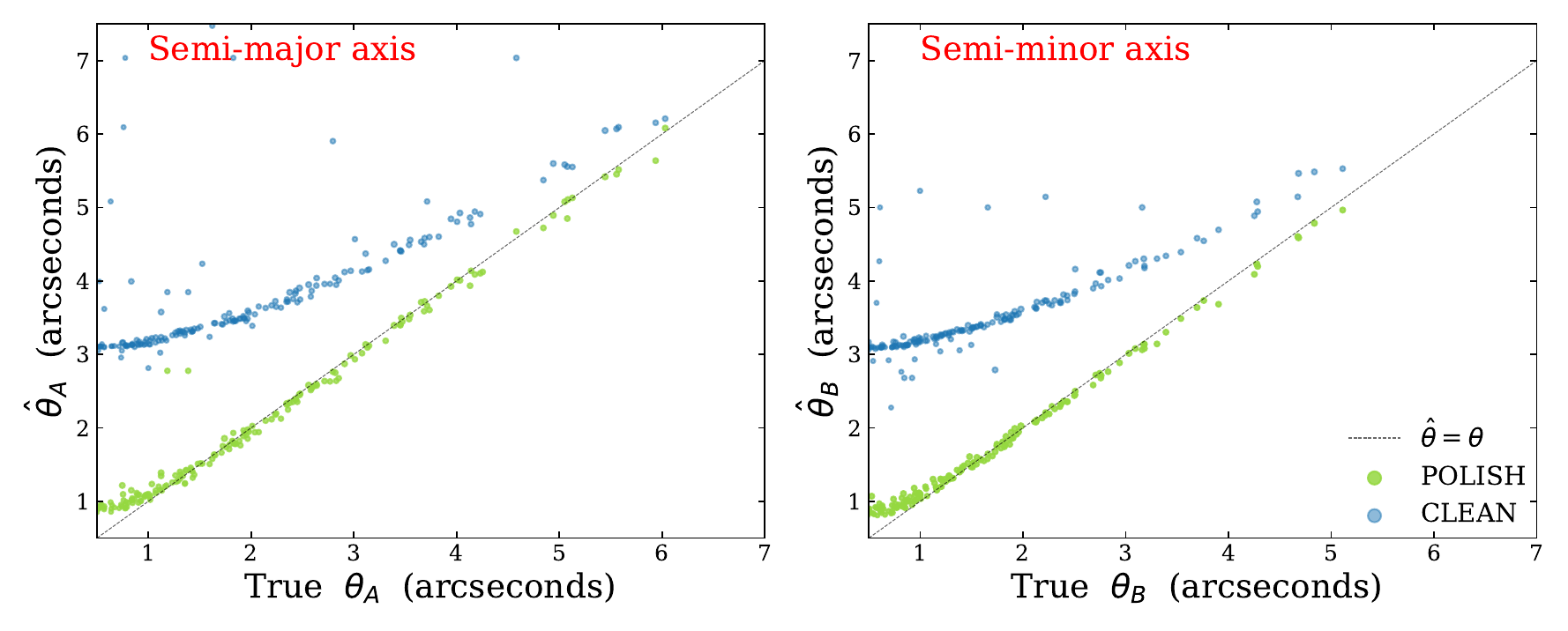}
  \caption{A comparison of the resolving power of POLISH (green dots) with that of traditional CLEAN algorithm (blue dots) on simulated DSA-2000 data with a 1300 MHz PSF kernel. 
  The vertical axes are the fitted semi-major (left) and semi-minor (right) axes, $\hat{\theta}$, of hundreds of ``galaxies" in the reconstructed images. They are plotted against the true galaxy sizes in arcseconds. CLEAN's resolving power asymptotes to the PSF angular scale preventing it from achieving super-resolution, while POLISH achieves accurate super-resolution well below the intrinsic resolution of the interferometer ($\approx3$\,arcseconds). Note that we have not pushed the limits of POLISH's 
  super-resolution because in this simulation the smallest 
  galaxies approach the pixel scale in size.
  }
 \label{fig:sr-clean-polish}
\end{figure}
\subsubsection{Super-resolution}

Super-resolution is critical for many scientific goals of the DSA-2000. We therefore investigate if spatial scales below the width of 
the PSF's central lobe ($\approx3$\,arcseconds) can be reliably
accessed with POLISH.
We quantify the super-resolution capability of POLISH using 
the galaxy simulations described in Section \ref{sec:simulation}.
First, the size of each ``galaxy" 
is extracted using the astronomical fitting software \texttt{Source-Extractor} \citep{sextractor}. This is done 
on the true image, CLEAN image, and the POLISH image.
The fitted galaxy sizes are then compared on a source-by-source basis; 
for each galaxy, we contrast the input (true sky) and output (POLISH and 
CLEAN reconstructions) semi-major ($\theta_A$) and semi-minor axes ($\theta_B$). If the reconstructed value, $\hat{\theta}$, is below the PSF angular scale of 
$\approx3$\,arcseconds, then  
super-resolution has been achieved. As can be seen in Fig.~\ref{fig:sr-clean-polish},
galaxies in the POLISH image are consistently the same 
size as galaxies in the true image, even at sub-arcsecond scales 
well below the DSA-2000's PSF size. In contrast, the traditional CLEAN algorithm 
cannot achieve super-resolution and instead 
asymptotes to the PSF 
central component's angular size. 

\subsubsection{Model mismatch}
\label{sec:psf-pert}

\begin{figure}[tb] 
 \centering
  \includegraphics[width=0.97\textwidth]{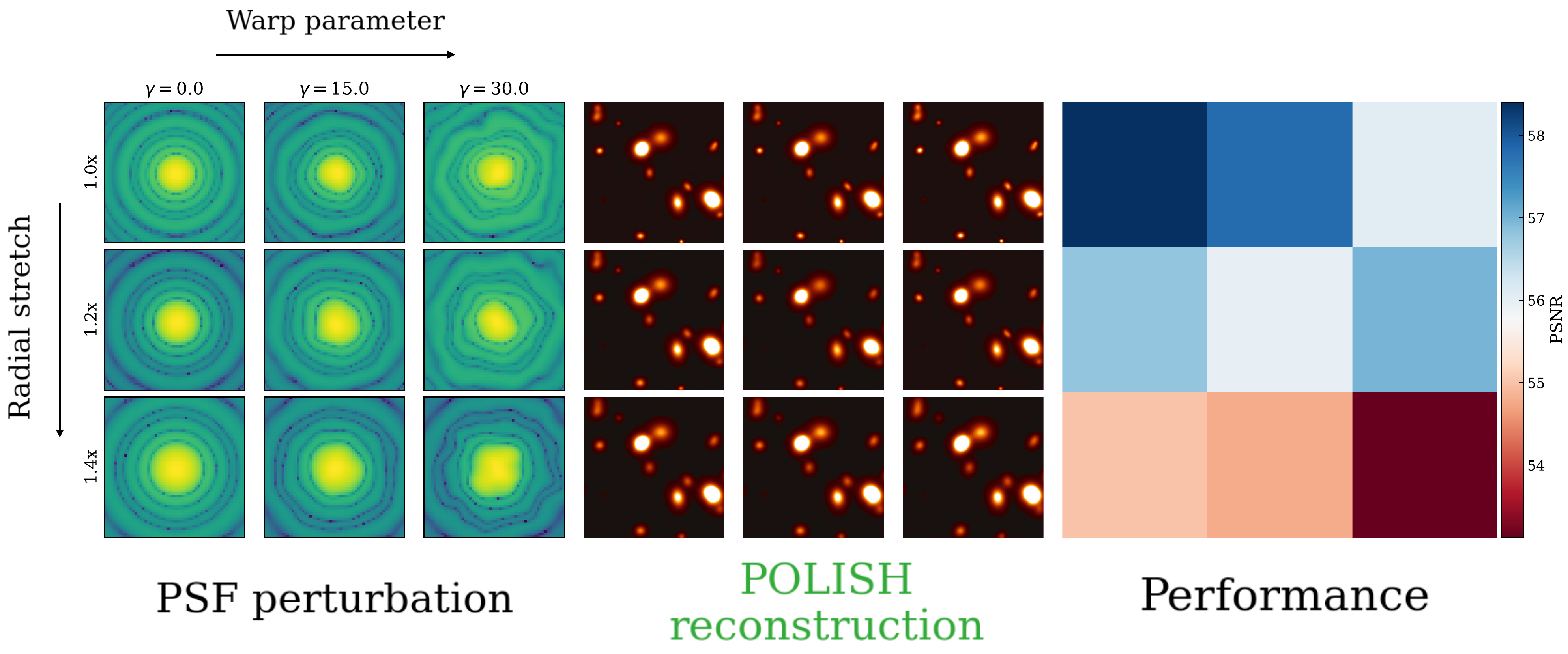}
  \caption{\small The impact of model 
  mismatch on reconstruction performance. POLISH has been trained using a distribution of PSF kernels with warps in the range of $\gamma$=0 to 20. We then validate on dirty images that have been corrupted by a PSF kernel that has been radially stretched (increasing downwards)
  and warped (increasing to the right), the logarithm 
  of which is shown on the left (zoomed in on the PSF's inner 
  28$\times$28\,arcseconds). The corresponding POLISH images and PSNRs are shown in the middle and right, respectively.}
 \label{fig:psf-pertubation}
\end{figure}

When a distribution of PSFs are used during training, 
POLISH can more robustly reconstruct the sky in the presence of instrumental calibration errors. We have performed an ablation study showing how performance decreases when training data is restricted to images corrupted with the ideal PSF kernel $k_{\mathrm{ideal}}$. 

In Fig.~\ref{fig:psf-pertubation} we demonstrate
 POLISH image reconstruction on test data whose PSF 
has been perturbed outside of the range of training data pertubations. 
In particular, POLISH has been trained with PSF kernels transformed by 
the warp function $\Lambda_2(\gamma)$ described in Section \ref{sect:PSF}, 
with $\gamma$ drawn from a uniform distribution between 0 and 20. The training 
data PSFs were not radially stretched for this figure, allowing us to 
test POLISH's performance under model mismatch.
As can be seen in the PSNR performance plot, the quality of the recovered test images degrades as the PSF perturbations get larger than the training set distortions. However, 
even in the relatively extreme distortions (bottom 
right element of each 3$\times$3 grid) PSNR remains 
above 50 (and comparable to or better than the typical CLEAN-based PSNRs above). Such distortions and uncertainty in the 
DSA-2000 PSF are greater than what we expect for actual experimental data; it is unlikely that the true PSF will be $40\%$ larger than 
the modelled PSF, and distortions as severe as the $\gamma=30$ ``warp'' would
require drastic mis-calibration.

\subsubsection{Deconvolving real astronomical data}
\label{sec:vla}

In practise, POLISH will be trained offline on simulated data and then applied to real data in quasi-real time. 
We show that POLISH
can reliably deconvolve real observations, having only been trained 
on simulations.
To demonstrate this, real astronomical 
observations \citep{hallinan-2017} were obtained from the VLA, a Y-shaped, 27-telescope interferometer in New Mexico, including a modelled PSF for that 
observation's configuration. We then trained 
POLISH on our radio galaxy simulation described in 
Sect.~\ref{sec:simulation}, but with $I_d$/$I_{sky}$ image 
pairs created with the VLA PSF.

\begin{figure}[tb]
 \centering
  \includegraphics[width=0.99\textwidth]{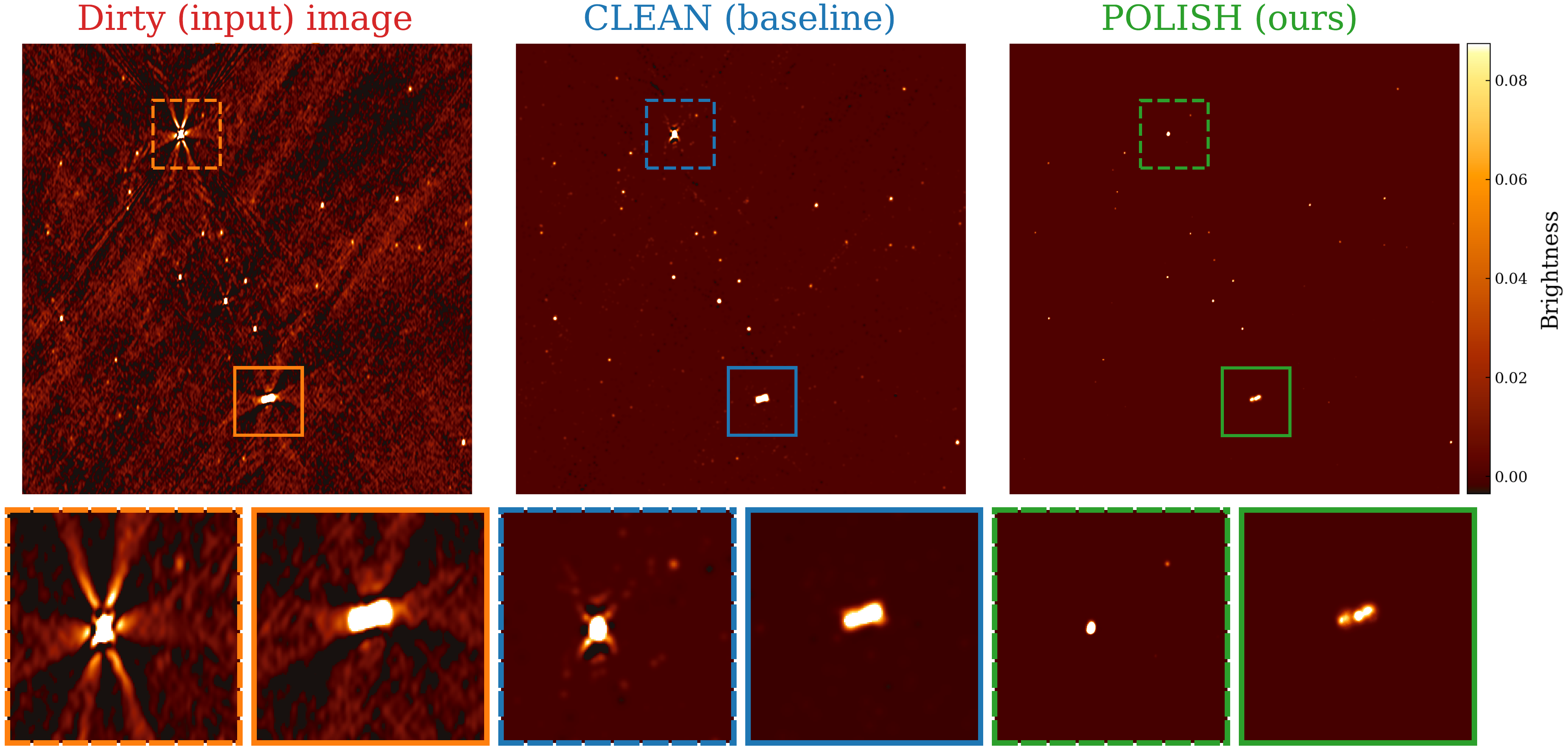}
  \caption{An example of POLISH being applied to real astronomical data from the Very Large Array (VLA), having been 
  trained only on simulated data. The left panel shows the 
  ``dirty image'' as measured by the VLA, and the right two panels show the traditional CLEAN algorithm's (middle) and our proposed approach POLISH (right). Specifically, these are the image reconstructions obtained by the two distinct deconvolution 
  algorithms.}
 \label{fig:vla}
\end{figure}

Reconstructing the sky from VLA data is a more challenging problem than 
what we will be faced with at DSA-2000, as it has far fewer antennas 
and therefore a much more degraded PSF (Fig.~\ref{fig:NNarchitecture}). Evidenced by the 
strong artifacts in the dirty image of Fig.~\ref{fig:vla}, the VLA's 
PSF is much less suppressed off-axis than that of DSA-2000. The PSF is also
highly azimuthally asymmetric due to the Y-shape of the array, leading to  
six radial ``struts'' expanding outwards from each source in the dirty image. 

In Fig.~\ref{fig:vla} we demonstrate that 
POLISH can deconvolve the real VLA dirty image, leaving 
behind fewer imaging artifacts than the traditional image-plane CLEAN algorithm. 
Fig.~\ref{fig:vla} shows a comparison of the dirty (input) image, 
CLEAN reconstruction, and POLISH reconstruction (left to right) for a 0.1$\times$0.1\,$\deg$ region 
of the 0.4$\times$0.4\,$\deg$ full image, including a further zoom in on particularly 
bright radio galaxies (bottom row). The difference between 
CLEAN and POLISH for the bright galaxies is striking. 
CLEAN leaves behind residual structures from the PSF, whereas POLISH produces 
shapes like those 
expected for distant radio galaxies. These residual structures include speckles that are likely not real 
radio point sources, as they lie in a line on 
the struts of the PSF response from bright galaxies; 
this is consistent with our the increased 
false-positive rate of CLEAN vs. POLISH 
in our simulated sky reconstruction.

We have chosen not to compare 
POLISH with RML methods for the following reasons. There exists a wide spectrum of 
RML algorithms and a true comparison would require comparing POLISH against a large set of reconstructions, each with its own finely tuned 
set of regularization parameters; some would perform poorly, some may do well, depending on our choice of parameters. We 
therefore cannot construct an apples-to-apples 
comparison that would meaningfully contrast an efficient, feed-forward algorithm such as POLISH with RML methods.

This result is very promising for two reasons.
First, we have shown that a relatively simple simulated data set can train a neural network 
that will deconvolve real radio inteferometric data. Second, the VLA's sampling 
of the Fourier plane is much less complete than that of DSA-2000.
The fact that POLISH can still perform under these circumstances 
shows that deep learning based techniques are not only a promising solution for the DSA-2000, but also for other current and future radio interferometers \citep{meerkat,askap,ska}.

\section{Radio weak lensing \& DSA-2000}
\label{sec:weaklensing}
Weak gravitational lensing is a powerful probe of the Universe's 
composition and expansion history, and is an essential tool for 
constraining the physical nature of dark energy \citep{kaiser-1996, hoekstra, mandelbaum2018weak}. 
Weak lensing imparts a distortion on distant 
galaxy shapes known as ``shear'', causing a separation-dependent correlation 
between galaxy shapes that is determined by the Universe's large scale structure 
and its evolution with time. Measuring the power spectrum of this cosmic shear allows 
us to constrain the cosmological parameters. To date, weak lensing research has been significantly more advanced at optical wavelengths. However, the achromaticity of gravitational lensing 
means the large scale structure will imprint itself 
the same way across the electromagnetic spectrum, while the systematics 
across wavelengths will vary. Observations in different bands are therefore highly complementary. For a detailed treatment of radio weak lensing in 
the context of the Square Kilometer Array (SKA) see 
\citep{brown-ska-2015,ska-lensing-I, bonaldi-ska-II, ska-III}. 

In the radio, few surveys thus far have had 
sufficient angular resolution, sensitivity, and sky coverage 
to detect the large numbers of SFGs required 
to do weak lensing cosmology. Using $\sim$\,9$\times10^5$ sources 
in the FIRST survey at 1.4\,GHz, \citet{chang-2004} fit galaxies 
in UV space to make the first-ever radio detection of weak lensing. \citet{Demetroullas-2016} cross-correlated an optical shear map 
containing $\sim$\,9 million galaxies in the Sloan Digital Sky Survey (SDSS) with a radio shear map from FIRST, resulting in a 2.7\,$\sigma$ detection of weak lensing. 
More recently, the SuperCLASS survey used 
0.26\,deg$^2$ of radio data from eMERLIN and the VLA, as well as 1.53\,deg$^2$ of optical data from the Subaru telescope, but had too few sources 
to make a detection \citep{harrison-superclass}. 
Despite the required catch-up at long wavelengths, there 
are benefits to doing weak lensing in the radio.
For example, the PSF of 
a radio interferometer can be known a priori through 
forward modelling, 
which is not true for optical telescopes. Furthermore, 
as both optical and radio lensing experiments will likely be 
systematics limited, cross-correlation provides a robust 
and unbiased estimator of the shear powerspectrum \citep{ska-III, harrison-superclass}.
And finally, sensitive radio telescopes such as DSA-2000 and the 
SKA are expected to probe higher redshifts 
than comparable optical surveys, with a fatter tail at high $z$.

The DSA-2000 will survey the whole sky north of declination $-30^\circ$
between 0.7 and 2.0\,GHz
down to a root-mean square (RMS) noise of just 0.5\,$\mu$Jy. 
The survey will contain approximately $10^9$ sources, primarily star forming radio galaxies. 
This is over one thousand times more galaxies 
than the FIRST sample where a radio weak lensing signal was first detected \citep{chang-2004}, and with a much cleaner PSF than the VLA (see Fig.~\ref{fig:NNarchitecture}). 
However, with a 
point spread function (PSF) whose full-width at half maximum (FWHM) is roughly 3'', many radio galaxies will be unresolved for the DSA-2000 in the dirty image. Using POLISH 
to achieve super-resolution, DSA-2000 could be competitive and complementary with Stage IV weak lensing 
experiments such as Euclid and the Vera C. Rubin Observatory. Shear maps from DSA-2000 will therefore enable a 
symbioses between wavelengths, where cross-correlating radio/optical shear maps allow for tight 
constraints on both dark energy and modified gravity \citep{ska-lensing-I}. 

\begin{figure}[htbp] 
 \centering
  \includegraphics[width=0.5\textwidth]{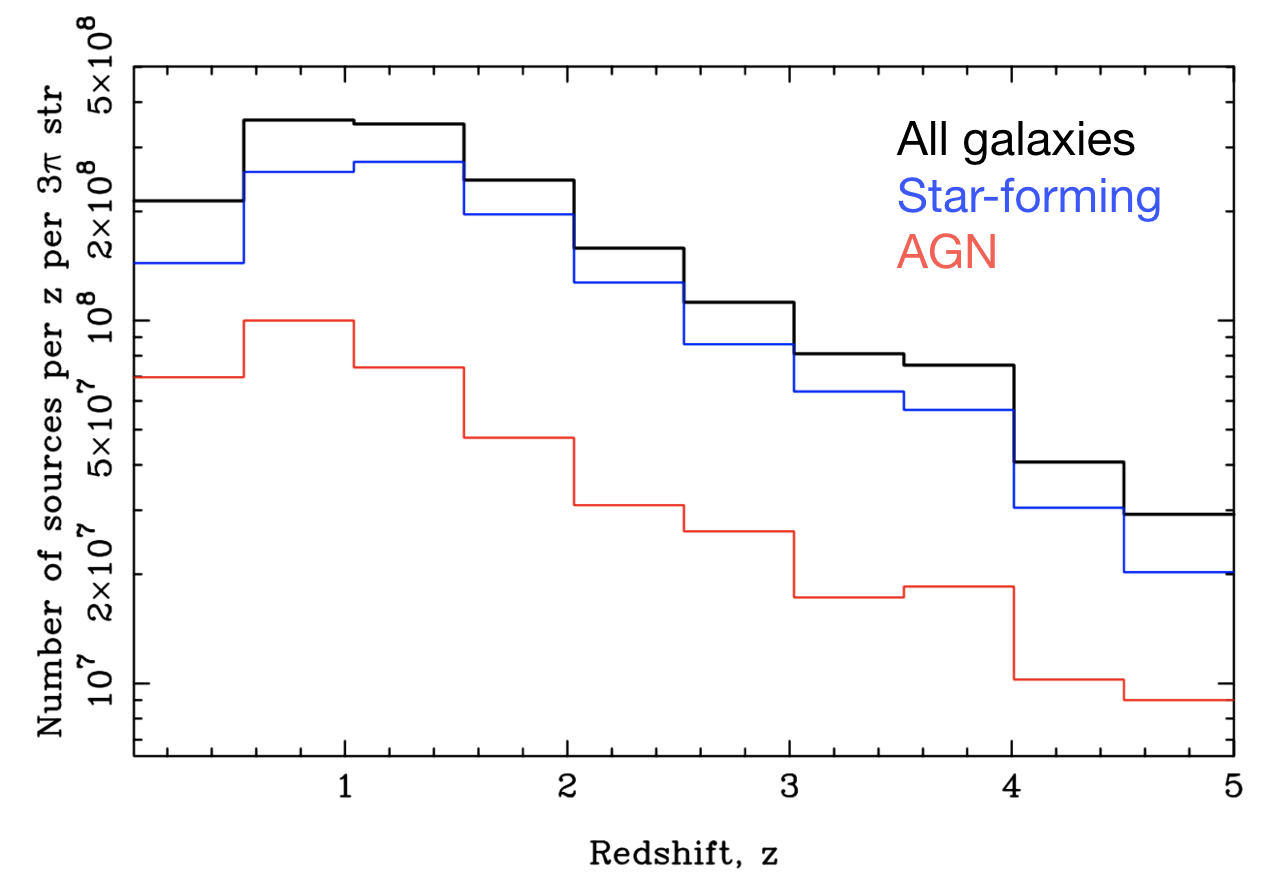}
  \caption{\small DSA-2000 
           will detect an enormous number of star forming galaxies.
           These can be used for weak lensing cosmology if 
           resolution below the telescope's PSF width can 
           be achieved. }
 \label{fig:nofz}
\end{figure}

\subsection{Radio weak lensing systematics}

Shape noise on the shear powerspectrum falls in proportion to 
the number of galaxies in the survey. 
But systematic uncertainties 
will play a large role in upcoming weak lensing surveys, whether in the 
radio or optical. For example,
the intrinsic alignment of galaxies can break the assumption 
that galaxy position angles and shapes are inherently uncorrelated. Biases in the measurement of shear must be accounted 
for and carefully calibrated out, as well as uncertainty in 
redshift estimation \citep{mandelbaum2018weak}. 
Finally, the PSF must be deconvolved out of 
the image, and residual PSF errors will lead to 
galaxy shape distortions. In the case of ground-based optical 
telescopes, time-variable seeing and related systematics 
make PSF modelling a central effort in weak lensing analysis \citep{gatti-des-shapes,jarvis-des-psf}. 

In the radio, weak lensing experiments suffer from 
overlapping but distinct systematics. Intrinsic alignment bias will affect 
galaxies in any weak lensing survey, 
but can potentially be calibrated out in the radio using the
polarized emission of star forming galaxies, which 
is unaltered by lensing \citep{brown-2011, ska-lensing-I}.
The PSF of a radio interferometer can be modelled deterministically 
(see Section~\ref{sec:imaging}), 
in contrast to an optical telescope. But PSF 
modelling is still a source of significant uncertainty because the 
PSF of an interferometer is typically poorly behaved, with significant 
sidelobes and azimuthal structure thanks to sparse sampling 
of the aperture.
This makes image reconstruction, and therefore shear measurement, 
more sensitive to errors in the PSF model even if the 
response can be forward modelled. 

The DSA-2000 will skirt some of these issues by densely 
sampling the UV plane and by optimizing antenna 
configuration to produce an azimuthally smooth PSF and 
highly attenuated sidelobes. While its suppressed sidelobes 
and smoothness will reduce errors introduced by PSF modelling 
and deconvolution, the nominal beam size of
3'' means that many SFGs will have angular size below our diffraction limit, hence our requirement of super-resolution. The DSA-2000 will therefore have fewer image reconstruction relics 
that bias galaxy shape measurement, but will be limited by 
resolution. For reference, the Dark Energy Survey has a median seeing of 
roughly 1'' and has produced strong cosmological constraints from its 
weak lensing experiment \citep{des-shapes-y3}, which sets an achievable 
baseline for POLISH.

\subsection{Forecasting}
We would like to estimate how well a weak-lensing survey on 
DSA-2000 will be able to constrain the cosmological parameters, 
both in the case where super-resolution of a factor of a 
few is achieved (POLISH SR in Table~\ref{tab:survey-params}) and when spatial resolution is limited by the synthesised beam size of 3'' (No SR in Table~\ref{tab:survey-params}). We compare simulations for DSA-2000 with simulations of 
SKA1 and the Euclid. Euclid is a visible to near-infrared space telescope 
expected to launch in mid- to late-2022 and is considered 
a Stage IV experiment. For these forecasts we use the same parameters 
assumed by \citet{ska-lensing-I}. SKA1 will be a mixed 
array of 133 15\,m SKA dishes and 64 13.5\,m dishes from the MeerKAT telescope\footnote{https://www.skatelescope.org/wp-content/uploads/2021/02/22380\_SKA\_Project-Summary\_v4\_single-pages.pdf} and a nominal 
PSF of 0.5''. Its weak lensing survey will be on Band 2 at 
0.95--1.76\,GHz \citep{ska-lensing-I}. As an empirical 
benchmark, we take the recently-published Dark Energy Survey (DES) 
Year 3 Results \citep{des-3year}. 
DES is a Stage III experiment that is already 
producing cosmological strong constraints from its optical weak lensing 
survey. We show these survey parameters in Table~\ref{tab:survey-params}.
\begin{table}[]
\center
\begin{tabular}{cccccc}
\hline
Survey          & $N_{gal}$       &     Sky coverage (deg$^2$)    &  $n_{gal}$ (sq arcmin$^{-1}$)                    &  $f_z$                  &    $\sigma_{pz}$    \\ \hline
\multicolumn{1}{c}{DSA-2000: POLISH SR} & \multicolumn{1}{c}{10$^9$} & \multicolumn{1}{c}{32,000} & \multicolumn{1}{c}{6/3$^\dagger$} & \multicolumn{1}{c}{0.3} & \multicolumn{1}{c}{0.05} \\
\multicolumn{1}{c}{DSA-2000: No SR} & \multicolumn{1}{c}{10$^9$} & \multicolumn{1}{c}{32,000} & \multicolumn{1}{c}{6/0.5$^\dagger$} & \multicolumn{1}{c}{0.3} & \multicolumn{1}{c}{0.05} \\
\multicolumn{1}{c}{SKA1}   & \multicolumn{1}{c}{5$\times10^7$} & \multicolumn{1}{c}{5,000} & \multicolumn{1}{c}{2.7} & \multicolumn{1}{c}{0.15} & \multicolumn{1}{c}{0.05}  \\ 
\multicolumn{1}{c}{Euclid}   & \multicolumn{1}{c}{1.6$\times10^9$} & \multicolumn{1}{c}{15,000} & \multicolumn{1}{c}{30} & \multicolumn{1}{c}{0.33} 
& \multicolumn{1}{c}{0.04} \\ 
\multicolumn{1}{c}{DES Y3}  & \multicolumn{1}{c}{$10^8$} & \multicolumn{1}{c}{4,143} & \multicolumn{1}{c}{6.7/5.6$^\dagger$} & \multicolumn{1}{c}{0.0}  & \multicolumn{1}{c}{0.05}  \\ 
\hline
\end{tabular}

\caption{Parameters used for forecasting the constraining power of a weak lensing 
experiment on different surveys, including DSA-2000 with and without POLISH's super-resolution. $N_{gal}$ is the total number of detections, 
$n_{gal}$ is the on-sky density, $f_z$ is the fraction of galaxies 
with spectroscopic redshift, and $\sigma_{pz}$ is the uncertainty 
on photometric redshifts. 
The DES Y3 row corresponds to the real values obtained 
from the year 3 galaxy shape catalog \citep{des-shapes-y3}; the other rows are estimates of survey parameters. \\
$\dagger$ Effective aerial number density of galaxies. On DSA-2000 we assume that only 
half of galaxies will be sufficiently super-resolved to produce 
a shape measurement.}
\label{tab:survey-params}
\end{table}
We base our estimate of the effective number of usable galaxies 
on both the estimated PSF sizes and the observed distribution of 
galaxy sizes. \citet{owen-2008} find that 
half of SFGs above 15\,$\mu$Jy have major-axes greater than 1''. However, recent work has argued that
$\mu$Jy radio sources may be more compact
than previously believed, with a median FWHM of 0.3'' \citep{cotton-2018, jimenez-andrade-2021}. 
For this reason, we assume that only one half of sources will be resolved with POLISH, assuming an effective resolution of 3''/SNR. For the case where super-resolution 
is not achieved we assume under 10$\%$ of sources will be resolved. For this reason 
we take the effective areal density for DSA-2000 with POLISH to be 3 galaxies per square arcminute, while 
only 0.5 galaxies per square arcminute will be useable without 
super-resolution. We assume that SKA1 will be able to 
resolve $\mathcal{O}(1)$ of detected galaxies, so we do 
not adjust its $n_{gal}$. The same is true for Euclid, which is 
space-based and therefore diffraction limited at its nominal 
resolution of 0.2''.

Forecasting weak lensing constraints requires each survey's
expected redshift distribution, areal density of galaxies, $n_{gal}$, 
total sky coverage, and information 
about the error in lensed galaxy redshifts. Obtaining redshifts of the 
lensed galaxies is an important component of weak lensing, 
as the cross-correlation angular powerspectra
between different redshift bins offers information 
on the evolution of structure over cosmic time \citep{bonnet-2016}.
For this analysis, we follow \citet{ska-lensing-I}
in using the MCMC parameter estimation code 
CosmoSIS \citep{zuntz-2015}. Specifically, we 
use code they have developed for forecasting the 
SKA's radio weak lensing survey\footnote{https://bitbucket.org/joezuntz/cosmosis-ska-forecasting/src/master/}, but with the putative DSA-2000 parameters \citep{dsa-2000-whitepaper}. 
On DSA-2000, we assume 15$\%$ of source will have spectroscopic redshifts (i.e. zero redshift uncertainty). 
The error 
on photometric redshifts will be 
approximately 0.05. DSA-2000 will obtain galaxy redshifts using several 
different methods. Approximately 30 million galaxies at $z<1$ 
will have spectroscopic redshifts from the DSA's HI survey. The SPHEREx 
near-infrared all-sky survey \citep{spherex-cosmology} will determine the redshifts of over one billion galaxies with $\sigma_z / (1+z) \leq 10\%$ and a subset of 
several hundred million galaxies with $\sigma_z / (1+z) \leq 1\%$ (see here\footnote{https://github.com/SPHEREx/Public-products} for current forecasts). Of these, three quarters can be cross-matched 
with the DSA-2000 catalog.

\begin{figure}[htbp] 
 \centering
  \includegraphics[width=0.95\textwidth]{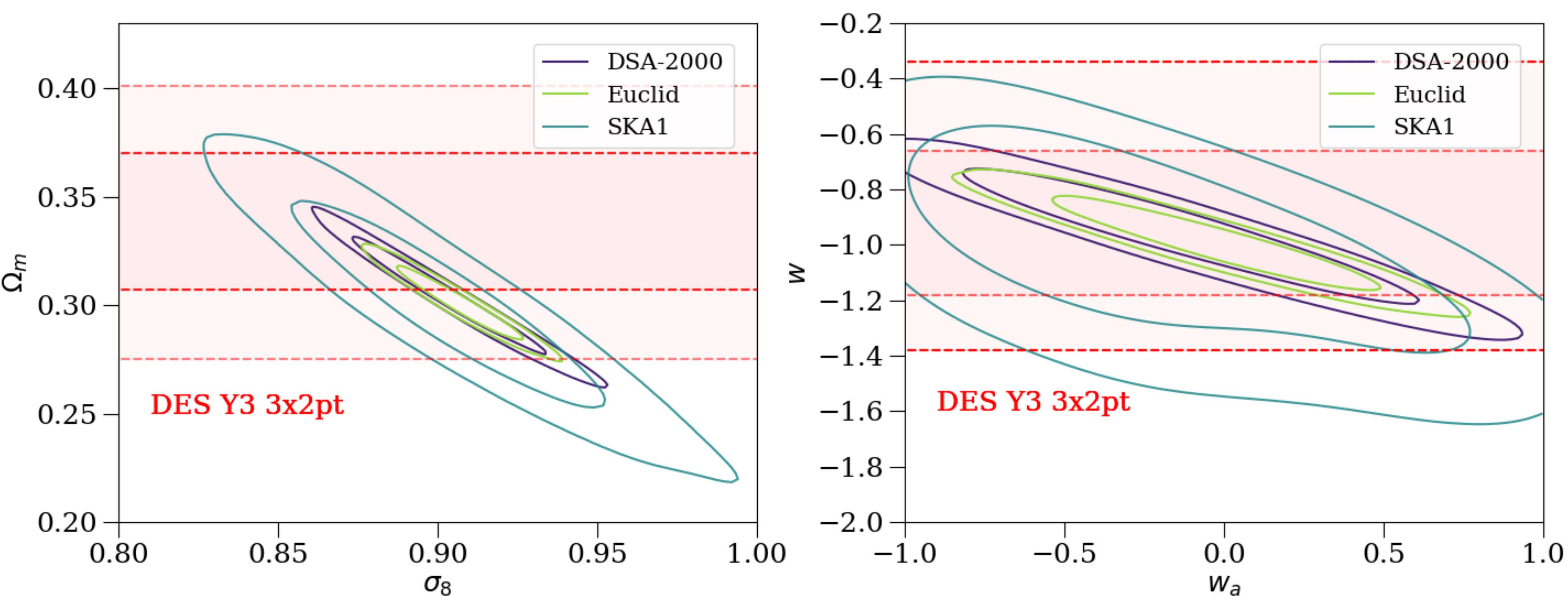}
  \caption{\small Marginalized constraints 
  on dark energy equation of state parameters, $w$ and $w_a$ (left), 
  and matter powerspectrum parameters, $\Omega_m$ and $\sigma_8$ (right) 
  for weak lensing surveys on DSA-2000 (with super-resolution), Euclid, and SKA1. 
  The red shaded regions are the 
  1- and 2-$\sigma$ constraints from the Dark Energy Survey Year 3 (DES Y3 3x2pt) 
  data after combining three two-point correlation functions (cosmic shear, 
  galaxy clustering, and shear with lens galaxy positions) \citep{des-3year}. Note 
  that the forecasts for DSA-2000, Euclid, and SKA1 are cosmic 
  shear only, but do not include lensing systematics.}
 \label{fig:cosmosis}
\end{figure}

We focus here on DSA-2000's ability to constrain the matter density and matter power spectrum normalization, $\Omega_m$ and $\sigma_8$ respectively, 
as well as the dark energy equation of state parameters $w$ and $w_a$.
In Fig.~\ref{fig:cosmosis} we show the cosmological parameter posteriors produced by CosmoSIS for 
DSA-2000, Euclid, and SKA1. The red regions 
show the 1- and 2-$\sigma$ constraints from DES Y3 results. We use their three two-point (3x2pt) 
correlation 
functions in cosmic shear, galaxy clustering, and shear with lens galaxy positions \citep{des-3year}. We find that SKA1's weak lensing survey 
will be comparable to the DES 3x2pt constraints in constraining 
both the matter powerspectrum and dark energy. DSA-2000 
produces constraints that are nearly as tight as Euclid. 
All four surveys provide an excellent opportunity for cross-correlation
and cross-validation, particularly between the different wavelengths \citep{ska-lensing-I, ska-III}.

\begin{figure}[htbp] 
 \centering
  \includegraphics[width=0.95\textwidth]{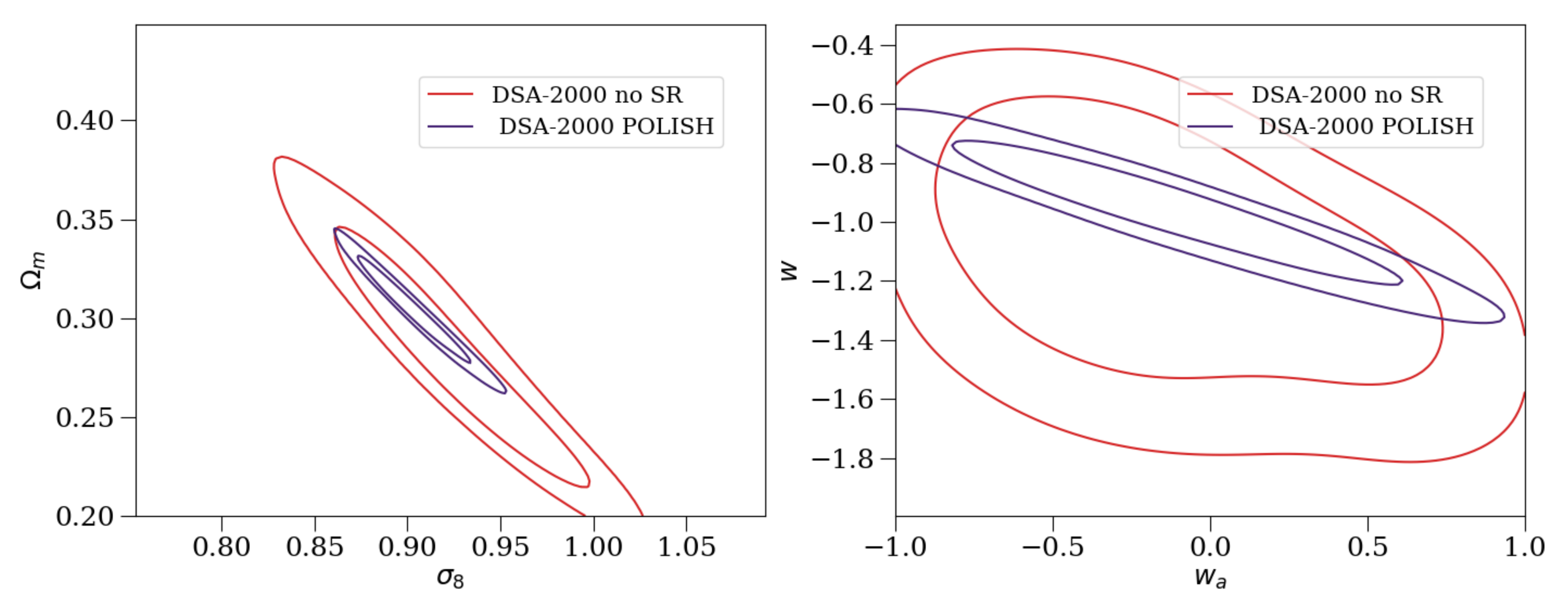}
  \caption{\small Comparing the ability of weak lensing survey on 
  DSA-2000 to constrain the cosmological parameters 
  with (purple) and without (red) POLISH's super-resolution.}
 \label{fig:cosmosis-SR}
\end{figure}

\section{Discussion \& conclusion}
\label{sec:discussion}

We have developed and demonstrated POLISH: a high dynamic range residual neural network for fast, accurate and super-resolved astronomical imaging with radio interferometers. Feed forward approaches like POLISH will be essential in analyzing data from upcoming instruments with enormous data rates where stream processing is critical, such as the upcoming DSA-2000 \citep{dsa-2000-whitepaper}. 
We have shown 
that POLISH is highly effective 
at deconvolving the PSF in both simulated and real astronomical data. It also achieves 
super-resolution and outperforms the
interferometric deconvolution algorithm of the last five decades, 
image-plane CLEAN, in both reconstruction accuracy (PSNR and SSIM) and resolution. We have shown how 
method forward-modelling approaches such as RML are not a 
viable option for surveys such as DSA-2000, where 
data rates preclude the preservation of visibilities. In its current form, POLISH (like CLEAN) does not provide 
uncertainty in image reconstruction. Looking forward, we plan 
to estimate uncertainty in the 
reconstructed image within the POLISH framework using 
Baysian deep learning to produce a posterior 
for $\hat{I}_{\rm sky}$.

Besides addressing the deconvolution problem in radio-interferometric images, POLISH makes possible new, currently inaccessible science by achieving super-resolution. When applied to data from the DSA-2000, POLISH will allow  shape measurements of nearly $\sim$\,10$^9$ galaxies, enabling a
weak lensing survey. We have forecasted the DSA-2000's 
ability to constrain cosmological parameters and compared it with both upcoming and current weak lensing experiments. We 
have also shown the significant improvement in 
constraints when achieving super-resolution with POLISH 
vs. when resolution is set by the diffraction limit.

For the specific application of 
weak lensing with POLISH, future work will benefit from a more detailed, forward modelled simulation of shear estimation systematics and PSF errors. 
Beyond the fraction of SFGs  
that can be resolved, it is important to know how different image reconstruction algorithms affect galaxy shapes as a function of S/N, galaxy size, redshift, and proximity to bright sources. This exercise would result in a mock galaxy shape catalog using one or more imaging techniques, including 
for other radio weak lensing experiments.
For example, the SKA will be less constrained by resolution than DSA-2000, but may benefit from the learning-based, accurate reconstructions provided by POLISH. The MeerKAT International GHz Tiered Extragalactic Exploration (MIGHTEE) survey will 
achieve $\mu$Jy sensitivity over 20\,deg$^2$ with a 
well-behaved PSF \citep{mightee}. Though its nominal 6'' resolution is larger than most SFGs, MIGHTEE would provide an interesting 
application of POLISH to radio weak lensing.
More generally, we anticipate that feed-forward techniques such as POLISH that deliver robust deconvolution and super-resolution will 
be valuable for upcoming high-volume radio surveys.

\section{Data availability}
The code for POLISH can be found at 
https://github.com/liamconnor/polish-pub/. This includes 
tools for simulating the microJansky radio sky as well as 
code to train and use the POLISH neural network.

\section*{Acknowledgements}
We are grateful to Schmidt Futures for 
supporting the Radio Camera Initiative, under which 
this work was carried out.
We thank Martin Krasser for his Tensorflow 2.x implementation of several super-resolution algorithms and his advice 
on this paper's application. We thank Joe Zuntz for help with his CosmoSIS code and Ian Harrison 
for valuable conversations.
We also thank Nitika Yadlapalli, Yuping Huang, 
and Jamie Bock for helpful discussion.

\bibliography{polish}{}
\bibliographystyle{aasjournal}



\end{document}